\documentclass{aa}
\usepackage{graphicx}
\usepackage{txfonts}
\newcommand{\hd}{HD~152246 }
\newcommand{\hde}{HD~152246}

\newcommand{\fotel}{{\tt FOTEL} }
\newcommand{\fotele}{{\tt FOTEL}}
\newcommand{\korel}{{\tt KOREL} }
\newcommand{\korele}{{\tt KOREL}}
\newcommand{\kms}{km~s$^{-1}$ }
\newcommand{\ks}{km~s$^{-1}$}
\newcommand{\vsin}{$v$~sin~$i$ }

\newcommand{\tef}{$T_{\rm eff}$ }

\newcommand{\lgg}{{\rm log}~$g$ }
\newcommand{\ms}{M$_{\odot}$}
\newcommand{\rs}{R$_{\odot}$}

\newcommand{\ubv}{\hbox{$U\!B{}V$}}
\newcommand{\bv}{\hbox{$B\!-\!V$}}

\newcommand{\m}{$.\!\!^{\rm m}$}
\begin{document}

   \title{HD\,152246 -- a new high-mass triple system\\ and its basic properties
\thanks{Based on data products from observations made with ESO telescopes at
La Silla Paranal Observatory under programmes
68.D-0095(A), 71.D-0369(A), 073.D-0609(A), 075.D-0061(A),
076.D0294(A), 077.D-0146(A), 079.D-0718(A), 081.D-2008(B),
083.D-0589(B), 086.D-0997(B), 087.D-0946(A), and 089.D-0975(A),
extracted from the ESO/ST-ECF Science Archive Facility,
and on the BESO spectra.}
}

   \titlerunning{A new triple system HD\,152246}

\author{
        A. Nasseri
        \inst{1}
        \and
        R. Chini
        \inst{1,2}
        \and
        P. Harmanec
        \inst{3}
        \and
        P. Mayer
        \inst{3}
        \and
        J.A. Nemravov\'{a}
        \inst{3}
        \and
        T. Dembsky
        \inst{1}
        \and
        H. Lehmann
        \inst{4}
        \and
        H. Sana
        \inst{5}
        \and
        J.-B.~Le~Bouquin
        \inst{6}
}
\institute{
                Astronomisches Institut,
                Ruhr--Universit\"at Bochum,
                Universit\"atsstr. 150,
                44801 Bochum, Germany
                \and
                Instituto de Astronom{\'i}a,
                Universidad Cat\'{o}lica del Norte,
                Avenida Angamos 0610,
                Casilla 1280
                Antofagasta, Chile
                \and
                Astronomical Institute of the Charles University,
                Faculty of Mathematics and Physics,\\
                V~Hole\v{s}ovi\v{c}k\'ach~2, CZ-180 00 Praha~8,
                Czech Republic
                \and
                Th\"uringer Landessternwarte Tautenburg,
                Germany
                \and
                ESA / Space Telescope Science Institute,
                3700 San Martin drive, MD 21218,
                Baltimore, USA
                \and
                Institut d'Astrophysique et de Plan\'etologie de
                Grenoble, CNRS-UJF UMR 5571, 414 rue de la Piscine,
                38400 St Martin d'He\`eres, France
                }

\authorrunning{A. Nasseri et al.}

\date{Received \today; accepted}


\abstract{Analyses of multi-epoch, high-resolution ($R \sim 50.000$) optical
spectra of the O-type star HD\,152246 (O9\,IV according to the most recent
classification), complemented by a limited number of earlier published radial
velocities, led to the finding that the object is a~hierarchical triple system,
where a close inner pair (Ba--Bb) with a slightly eccentric orbit ($e = 0.11$)
and a period of  6\fd0049 revolves in a 470-day highly eccentric orbit
($e = 0.865$) with another massive and brighter component A. The mass ratio of
the inner system must be low since we were unable to find any traces of
the secondary spectrum. The mass ratio A/(Ba+Bb) is 0.89. The outer system has
recently been resolved using long-baseline interferometry on three
occasions. The interferometry confirms the spectroscopic results and specifies
elements of the system. Our orbital solutions, including the combined
radial-velocity and interferometric solution indicate an orbital
inclination of the outer orbit of $112^\circ$ and stellar masses
of 20.4 and 22.8~\ms. We also disentangled the spectra of components A and Ba and compare them to synthetic spectra from two independent programmes, TLUSTY and
FASTWIND. In either case, the fit was not satisfactory and we postpone a better
determination of the system properties for a future study, after obtaining observations during the periastron passage of the outer orbit
(the nearest chance being March 2015). For the moment, we can only conclude
that component~A is an O9\,IV star with \vsin = $210\pm10$~\ks and effective
temperature of $33000\pm500$~K, while component Ba is an O9\,V object with
\vsin = $65\pm3$~\ks and \tef = $33600\pm600$.
}
\maketitle
%

\section{Introduction}

The only stable configuration of a stellar triple system seems
to consist of a close binary and a substantially more distant third
component.
These systems are important for testing theories of star formation and
stellar evolution in the presence of nearby companions.

The source HD~152246 (HIP~82685) is a member of the Sco OB1 association at a distance
of about 1.585~kpc \citep{Sung2013}. Although the star has been included
in many photometric and spectroscopic surveys of O stars, the knowledge of
its properties is quite limited.

The binary nature of \hd was discovered by \citet{Thack73} who published
six radial velocities (RVs) from 1963 -- 1967 showing a range of about 120 \ks,
but this finding has not received much attention.
\citet{Con77} published one new RV and noted that the star is a RV variable.
\citet{Penny96} suspected the presence of a weak secondary in the CCF peaks,
but it was only \citet{SL2001} who discovered the presence of sharp and
wide lines in the IUE spectra and published two RVs for both components.
They interpreted the two systems of lines as belonging to the primary and
secondary in a binary system.
\citet{Mason98} observed the star with speckle interferometry, but
could not detect any companion for the angular separation range
$0\farcs035 < \rho < 1\farcs5$ with $\Delta m < 3$.

The \hd system has been a~part of our monitoring programme for stellar multiplicity
\citep{Chini12} and attracted our attention because of its varying spectral
line profiles. In our preliminary report based on an analysis of the
\ion{He}{I}~5876\,\AA\ line \citep{Nass14} we showed that \hd must be
a~hierarchical triple system. What is actually observed is a combination
of a broad-line O-type star (component A) and another O-type star with
narrow lines, which itself is the primary component Ba of a close pair.
The RVs of the narrow-line star follow the orbital motion with an
invisible secondary Bb with a period of about 6 d and also a motion
around the common centre of gravity with the component A. The RVs of the
broad-lined component A follow only the orbital motion in the wide
orbit (we had tentatively derived a period of 53 days for it; we revise
the period here).

The combined spectral type of \hd was first determined by \citet{Morgan55}
as O9:\,III:. Later, more accurate but fairly similar properties were found:
\citet{Schild69} and \citet{Garr77} classified the star as O9\,III
while \citet{Wal73} and \citet{Thack73} assigned spectral types
of O9\,III-IV(n) and O8.5\,III, respectively. Several published
measurements of the visual brightness of \hd are mutually quite
consistent \citep{Thack73, KN77, Dachs82, Schild83, Hog00}. \citet{Mer98}
give the following weighted mean value $V = 7$\m308$\pm0$\m015.
The only deviating value was reported by \citet{HW84}: $V=8$\m09, which
we suspect it could be a misidentification of the star. We also used
all Hipparcos $H_{\rm p}$ observations with flags 0 and 1
\citep{ESA97} and transformed them to Johnson $V$ following \citet{Hec98}
to obtain $V=7$\m330$\pm$0\m008. The full range of individual values
is 0\m05, so a future test for micro-variability would perhaps be useful.
In any case it seems that no binary eclipses are observed.

Several -- seemingly contradictory -- determinations of the projected
rotational velocity of \hd were published (see Table~\ref{vsin}).
It seems clear now that \citet{CE77} probably derived \vsin from the
unresolved blend of lines of both stars, other investigators mainly
measured the width of the narrow component Ba until \citet{SL2001}
resolved and measured the lines of both spectra.

In this paper we present the analysis of 49 high resolution spectra,
obtain rather accurate orbital elements and estimate the
basic physical properties of the system.

\begin{figure}
\centering
\resizebox{\hsize}{!}{\includegraphics{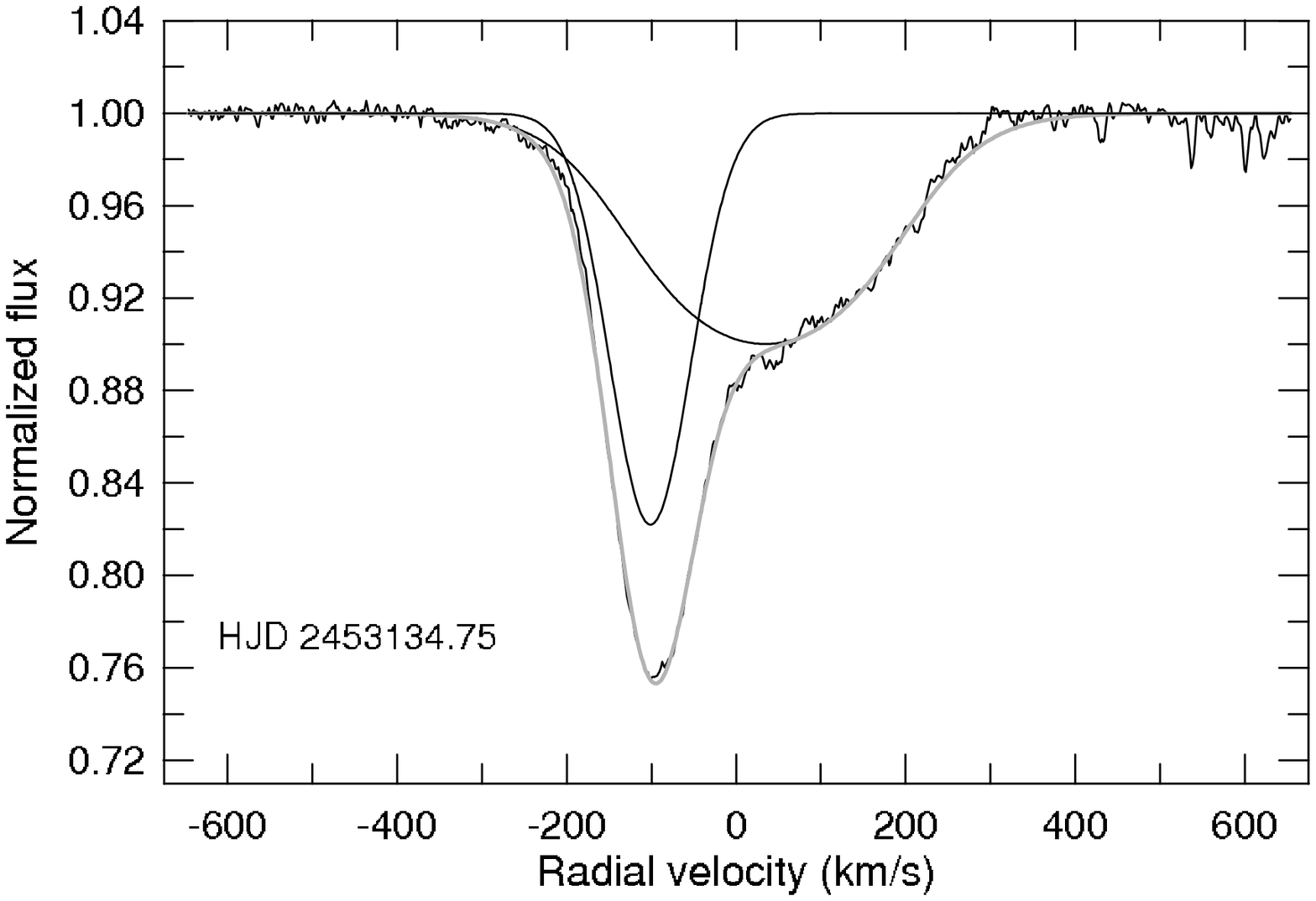}}
\resizebox{\hsize}{!}{\includegraphics{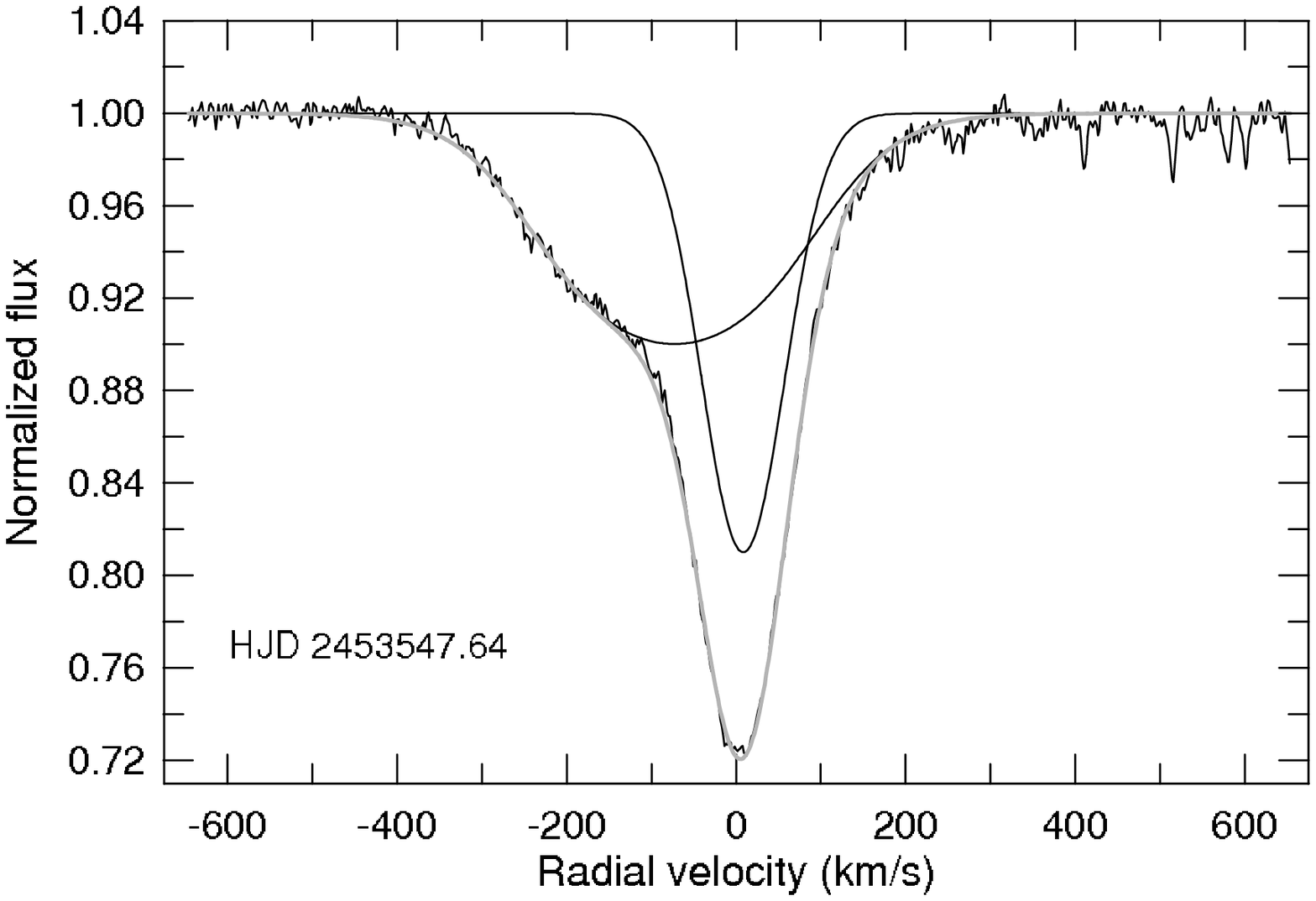}}
\caption{Two \ion{He}{I}~5876\,\AA\ line profiles from different orbital
 phases. The observed profiles have been fitted by a narrow Gaussian and
 a wide rotationally broadened profile. The resulting fits are denoted
 by grey curves.}
\label{profiles}
\end{figure}

\begin{table}
\caption{Previously published determinations of the projected rotational
velocity of \hde.}
\label{vsin}
\begin{tabular}{clcc}
\hline\hline\noalign{\smallskip}
\vsin (\ks) & Source   \\
\noalign{\smallskip}\hline\noalign{\smallskip}
280  & \citet{CE77}     \\
$90 \pm 20$ & \citet{Grigsby92} \\
103  & \citet{Penny96}  \\
72  & \cite{How97}   \\
59 \& 156 & \citet{SL2001} \\
\hline\noalign{\smallskip}
\end{tabular}
\end{table}

\begin{table}
\caption{Published and new RVs of components A and Ba. The new velocities are based on the fit of the \ion{He}{I}~5876~\AA\, line by a combination of a Gaussian and a rotationally broadened profile (see text for details).}
\label{rvtab}
\begin{tabular}{rrrlllc}
\hline\hline\noalign{\smallskip}
 RJD\ \ \ &RV$_{\rm A}$&RV$_{\rm Ba}$&Source\\
\noalign{\smallskip}\hline\noalign{\smallskip}
38251.2378&--&    36.0 & \citet{Thack73}\\
39200.6408&--&    60.0 & \citet{Thack73}\\
39200.6498&--&    52.0 & \citet{Thack73}\\
39688.3284&--&$  -18.0$& \citet{Thack73}\\
39720.2631&--&$  -58.0$& \citet{Thack73}\\
39723.2539&--&    50.0 & \citet{Thack73}\\
42888.78  &--&$  -29.8$& \citet{Con77}\\
44995.778 &$-109.4$&   46.6&\citet{SL2001}\\
45000.609 &$-103.0$&   26.5&\citet{SL2001}\\
\noalign{\smallskip}\hline\noalign{\smallskip}
51327.8888&$    -7.0$&$ -50.8$&this paper, FEROS\\
52040.8836&$   -49.0$&$ -20.5$&this paper, FEROS\\
52338.8394&$   -23.0$&    8.6 &this paper, FEROS\\
52383.7834&$   -25.0$&$ -44.3$&this paper, FEROS\\
52783.6903&$   -15.0$&    2.4 &this paper, FEROS\\
52784.6323&$   -19.0$&$ -28.5$&this paper, FEROS\\
53130.7748&     44.0 &$ -58.5$&this paper, FEROS\\
53131.6933&     37.0 &$ -39.6$&this paper, FEROS\\
53132.7522&     42.5 &$ -64.3$&this paper, FEROS\\
53133.7812&     37.0 &$ -94.0$&this paper, FEROS\\
53134.7494&     34.0 &$-101.1$&this paper, FEROS\\
53135.7069&     33.0 &$ -81.9$&this paper, FEROS\\
53547.6424&$   -73.0$&    8.3 &this paper, FEROS\\
53798.9178&$   -30.0$&    9.2 &this paper, FEROS\\
53862.8912&$   -41.0$&$  -1.4$&this paper, FEROS\\
53865.8991&$   -40.5$&$ -14.2$&this paper, FEROS\\
54212.8182&$   -26.0$&   12.4 &this paper, FEROS\\
54627.7027&$    -5.0$&$ -12.8$&this paper, FEROS\\
54976.7046&$   -42.0$&$  -7.2$&this paper, FEROS\\
55641.8195&$   -18.5$&   22.5 &this paper, FEROS\\
55642.8300&$   -19.0$&$  -9.4$&this paper, FEROS\\
55696.8448&$   -23.0$&$  -0.7$&this paper, FEROS\\
55697.7549&$   -22.0$&$ -35.6$&this paper, FEROS\\
55698.8283&$   -21.0$&$ -48.3$&this paper, FEROS\\
56068.7356&$   -17.5$&    3.9 &this paper, FEROS\\
56098.8047&$   -14.0$&    4.8 &this paper, FEROS\\
\noalign{\smallskip}\hline\noalign{\smallskip}
54913.8587&$   -48.0$&   20.3 &this paper, BESO\\
54920.8675&$   -68.0$&   46.3 &this paper, BESO\\
55050.5851&     18.5 &$ -82.1$&this paper, BESO\\
56485.5223&      1.5 &$ -78.8$&this paper, BESO\\
56490.6158&$    -1.0$&$ -72.8$&this paper, BESO\\
56495.6219&$    -4.0$&$ -31.9$&this paper, BESO\\
56498.7118&$   -10.0$&$ -45.9$&this paper, BESO\\
56502.5216&$   -10.0$&$ -65.2$&this paper, BESO\\
56508.4660&$    -3.5$&$ -62.4$&this paper, BESO\\
56512.5388&$    -3.5$&    9.9 &this paper, BESO\\
56516.5365&$   -12.0$&$ -48.3$&this paper, BESO\\
56518.6466&$   -10.5$&   11.4 &this paper, BESO\\
56521.5196&$   -13.0$&$ -71.1$&this paper, BESO\\
56532.5305&$   -24.0$&$ -59.7$&this paper, BESO\\
56540.5434&$   -18.5$&$ -43.1$&this paper, BESO\\
56541.5275&$   -13.0$&$  -4.0$&this paper, BESO\\
56542.5155&$   -13.0$&   15.4 &this paper, BESO\\
56545.4841&$   -12.0$&$ -62.6$&this paper, BESO\\
56546.5149&$   -13.0$&$ -39.5$&this paper, BESO\\
56547.5287&$   -21.0$&$  -0.8$&this paper, BESO\\
56548.5231&$   -20.5$&   17.1 &this paper, BESO\\
56576.4898&$   -20.0$&$ -43.7$&this paper, BESO\\
56585.4939&$   -20.0$&$  -5.5$&this paper, BESO\\
\hline\noalign{\smallskip}
\end{tabular}
\end{table}


\begin{figure*}
\centering
\resizebox{\hsize}{!}{\includegraphics{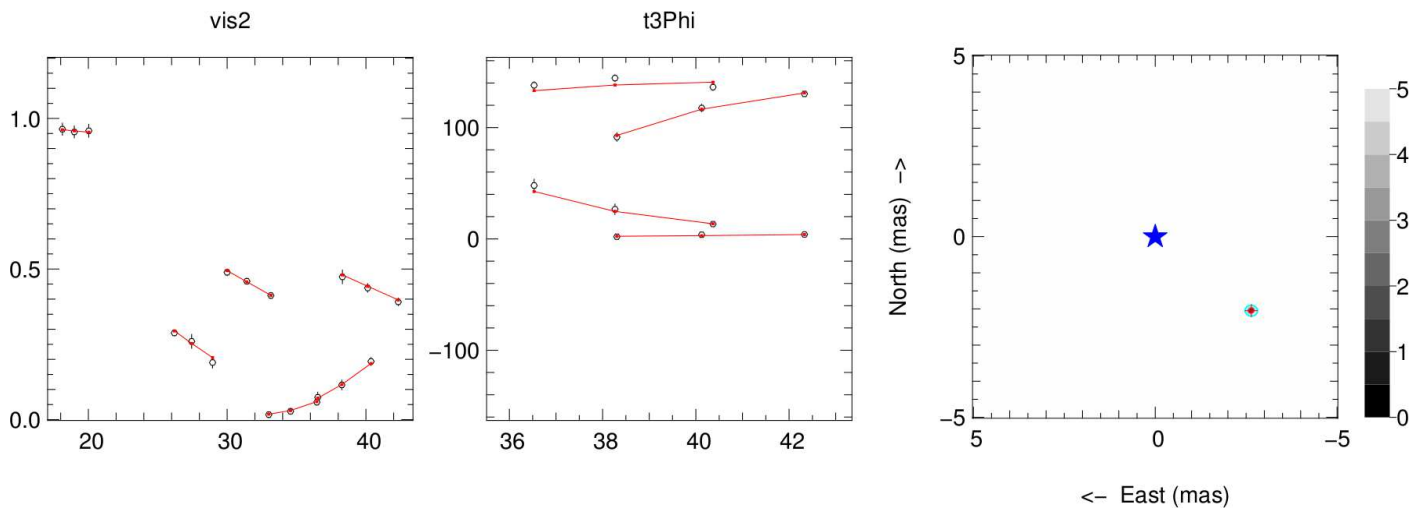}}
\caption{Visibilities (left panel), closure phase (middle panel) and
 $\chi^2$ map of the Feb 28 PIONIER observation of HD~152246.
 The best-fit binary model is overlaid.}
\label{fig: pionier}
\end{figure*}

\section{Observations and data reduction}
\subsection {Spectroscopy}
We have secured 49 high resolution ($R \sim 50.000$) echelle spectra
covering the wavelength range from 3620 to 8530~\AA\, and the time
interval from May 1999 to October 2013. Twenty-six spectra were taken at
the ESO with the spectrograph FEROS (RJDs 51327.9 - 56098.8); the first
two spectra from March and April 2002 stem from the ESO\,1.52\,m telescope,
the remaining data come from the MPG/ESO 2.2\,m telescope. Twenty-three
spectra were obtained with BESO (RJDs 54913.9 - 56585.5)
\footnote{Throughout this study we shall use the following abbreviation
for the {\sl reduced Julian date} RJD=HJD-2400000.0 \,\,.}
at the Universit\"atssternwarte Bochum on a side-hill of Cerro Armazones
in Chile. BESO \citep{Fuhrmann11} is a clone of the ESO spectrograph
FEROS on La Silla and has been attached to the 1.5\,m Hexapod-Telescope.
The identical spectrographs allowed a perfect combination of the two
data sets. All spectra were reduced with a pipeline based on a MIDAS
package adapted from FEROS. The signal-to-noise ratio of
the reduced FEROS spectra ranges from 247 to 487; that of the BESO
spectra from 65 to 177. For either instrument, the signal-to-noise
ratio was measured in the neighbourhood of the \ion{He}~{I} line at
5876~\AA.

At a later stage of the analysis, we also used 9 published RVs of
component B and provided a few comments on 2 published RVs of
component~A (Table~\ref{rvtab}).

\begin{table*}
\centering
\caption{Interferometric best-fit measurements and 1$\sigma$\ error-bars.}             
\label{tab: pionier}
\begin{tabular}{l l l l l }
\hline\hline  & & \multicolumn{2}{c}{Observing date} \\
Parameter & Unit & 2014-Feb-24 & 2014-Apr-04 & 2014-May-08 \\
\hline
\vspace*{-3mm}\\
RJD                         &         & \hspace*{4mm} 56712.840         & \hspace*{4mm} 56751.778          & \hspace*{4mm} 56785.902        \\
$(f_B/f_{A})_{1.65\mu}$     &         & \hspace*{2.5mm}$ 0.77 \pm 0.01$ & \hspace*{2.5mm}$ 0.76 \pm 0.04$  & \hspace*{2.5mm}$ 0.79 \pm 0.1$ \\
$\delta$E                   & (mas)   &                $-2.64 \pm 0.17$ &                $-2.23 \pm 0.13$  &                $-1.72 \pm 0.19$\\
$\delta$N                   & (mas)   &                $-2.05 \pm 0.15$ &                $-2.23 \pm 0.07$  &                $-2.25 \pm 0.26$\\
$r$                         & (mas)   & \hspace*{2.5mm}$ 3.34 \pm 0.16$ & \hspace*{2.5mm}$ 3.15 \pm 0.10$  & \hspace*{2.5mm}$ 2.83 \pm 0.24$\\
$\theta$                    & (\degr) & \hspace*{0.7mm}$232.2 \pm 1.6 $ & \hspace*{0.7mm}$224.9 \pm 0.2 $  & \hspace*{0.7mm}$217.4 \pm 4.5 $\\
$\chi^2_\mathrm{red}$       &         & \hspace*{8mm}       0.6         & \hspace*{8mm}      0.9           & \hspace*{8mm}      0.8         \\
\vspace*{-3mm}\\
\hline                                   
\end{tabular}
\end{table*}

\subsection{Long-baseline interferometry}
Interferometric measurements of HD~152246 were obtained on February 24,
April 4 and May 8, 2014 using the PIONIER four-beam combiner \citep{LeBo}
at the four Auxiliary Telescopes of the ESO Very Large Telescope
Interferometer \citep[VLTI,][]{HAA08, HAB10}. The data were reduced with
the \texttt{pndrs} package \citep{LeBo}. Reduction steps and achieved
calibration accuracy on the O-type stars were described earlier
\citep{SLBM13} and will not be repeated here. The data analysis closely
follows the procedure used in the Southern Massive Star at High angular
resolution survey (Sana et al., submitted), which uniformly
analyses PIONIER observations of a sample of over 100 O-type stars.

HD~152246 was clearly resolved at three epochs with a separation of
about 3.2-3.3~milli-arcsec (mas). A clear rotation of the binary axis
could be detected over the 73-day baseline of our PIONIER observations.
Table~\ref{tab: pionier} provides the journal of the interferometric
observations together with the parameters of the best fit binary model.
Figure~\ref{fig: pionier} shows and example of the measured data and the best fit binary
model.


\begin{figure}
\centering
\resizebox{\hsize}{!}{\includegraphics{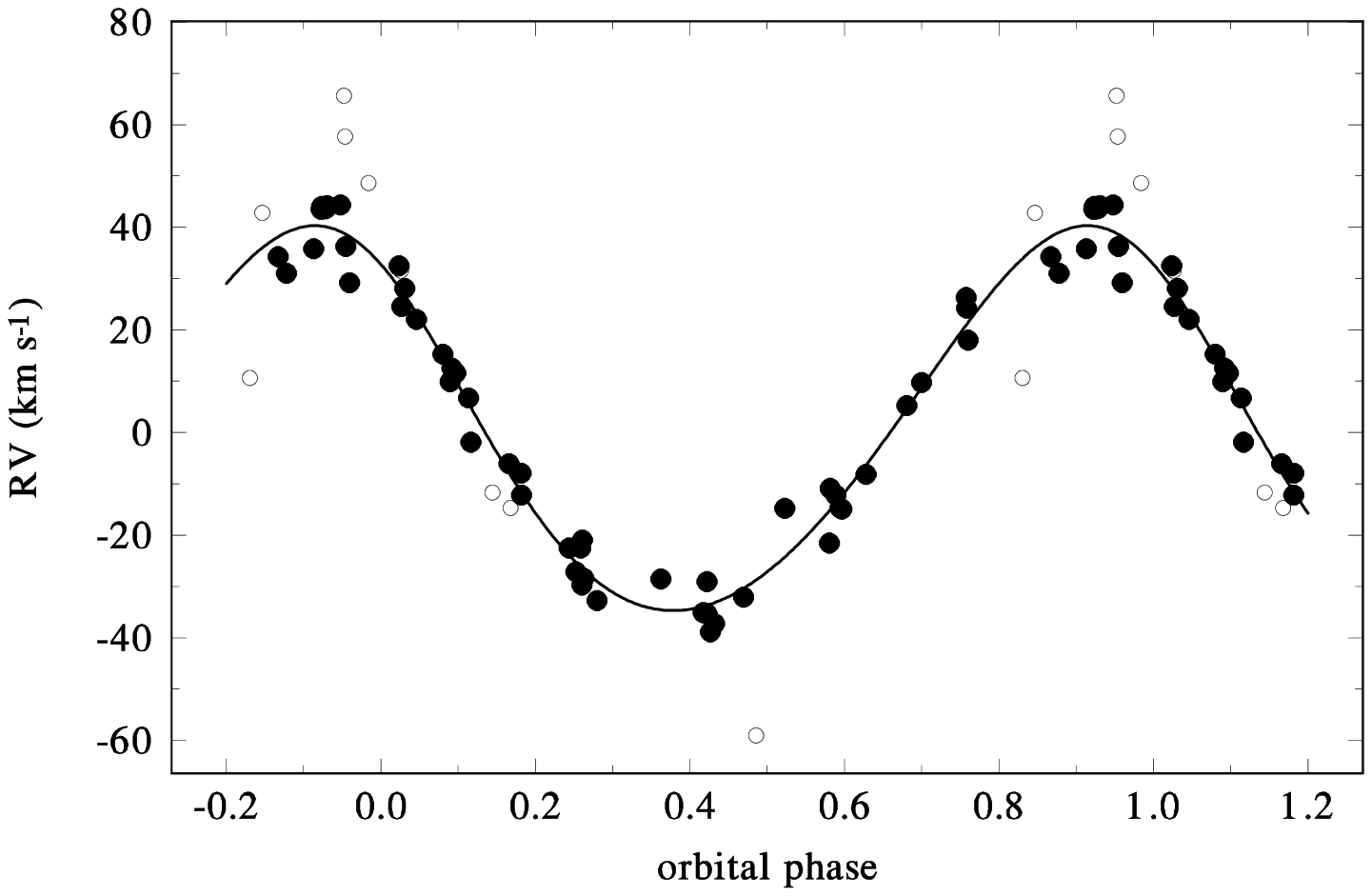}}
\resizebox{\hsize}{!}{\includegraphics{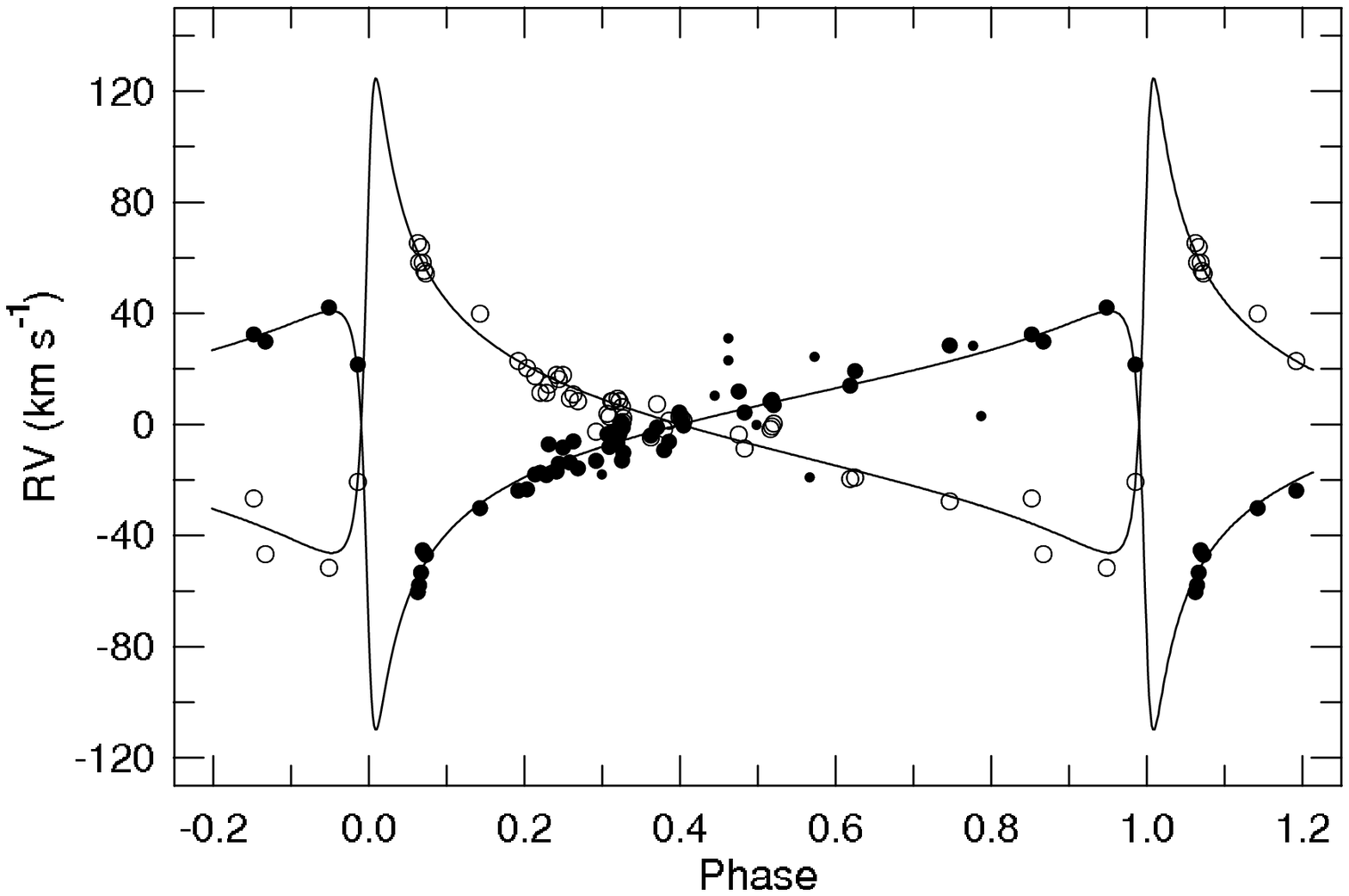}}
\caption{Phase plots for both binary orbits. {\sl Top panel:}
RVs of component Ba, prewhitened for the variations in the 470~d orbit
and plotted vs. phase for the 6 d period and periastron epoch
from Table~\ref{fotel}. Open circles denote RVs from the literature.
{\sl Bottom panel:} RV curves of components A and B in the wide orbit.
Open circles denote component A, filled circles component B (smaller ones
for RVs from the literature). We note that in both plots, RVs $-\gamma$ are
plotted.}
\label{orbits}
\end{figure}

\section{Line profiles, radial velocities and orbital periods}
To have a guidance to a more sophisticated analysis, we first measured
RVs of both components of the strong and usually well-exposed
\ion{He}{I}~5876\,\AA\ line, fitting the narrow component by a Gaussian
and the broader one by a rotationally broadened profile. The examples of
such fits are shown in Figure~\ref{profiles} and all RVs, together with
nine RVs from the literature, are in Table~\ref{rvtab}.

We note that \citet{Nass14} originally reported a tentative
value of 53 days for the longer period and 6 days for the shorter period.
Meanwhile we could recover a number of old spectra from the epoch 1999 - 2003
which led to a substantial revision of the originally suggested long period.

The analysis of the RV variations soon showed that the
RV of the component~A varies with a single period, which turned out to be
about 470~d. The RVs of component Ba are modulated also by this period
(showing anti-phase variations with respect to those of component~A), but
they also varied with a shorter period of about 6~d.  After deriving
a more accurate value of the short period, treating data subsets as having
individual systemic velocities (to allow for the changes with the 470~d
orbit), we included all RVs and derived an orbital solution for the triple system.
For this task the programme \fotel \citep[][and references therein]{Had04a} is
well suited. We also verified that the 6-d period is clearly seen
in the RVs for the narrow component published by \citet{Thack73}, \citet{Con77},
and \citet{SL2001}. The situation is a bit more complicated with the two
RVs for component~A as we shall discuss below.

In a second step, we therefore derived a solution for the triple-star
system with \fotele, using all Gaussian RVs for both components measured
in our new spectra and also all 9 RVs of component B from the literature,
first with equal weights and then weighting the three data subsets by the
weights inversely proportional to the square of root-mean-square
residuals (rms) from the preliminary solution. This solution is presented
in Table~\ref{fotel}. The radial-velocity curve for the 6 d orbit is
shown in Figure~\ref{orbits} and corresponds to Table~\ref{fotel}. The
curve for the 470 d orbit is presented in Figure~\ref{fig: 3Dorbit}; the
parameters based also on the interferometric results as listed in
Table~\ref{tab: 3Dorbit} were used for its construction.

\citet{SL2001} published two RVs for the component~A (see Table~\ref{rvtab}):

\smallskip
 $-$109.4 \kms at RJD~44995.7780, phase 0.7811, and

 $-$103.0 \kms at RJD~45000.6090, phase 0.7913

\smallskip\noindent
It is clear from Figure~3 (bottom panel) that these RVs
are for some $-50$~\kms too negative than our new RVs. The difference
could be made smaller if the true value of the long period would be
474.3~d. However, this change in the period would increase the rms of
the orbital solution of component B to 5.6~\ks, and even then, a
deviation of about 30~\kms would persist. One can conjecture that the
reason for too negative RVs measured by \cite{SL2001} lies in that our
results (see below) show that \vsin of component~A is about 200~\ks,
while they only derived 156~\ks. As the relation between the narrow and
broad profiles measured by \citet{SL2001} was similar to that in Figure~1
(bottom panel) here, the smaller \vsin means that the RVs of broad
profiles as measured by them might be erroneously shifted by about one
half of the \vsin difference, i.e. more than by 20 ~\ks. Such a shift
could explain the observed RV difference. The ultimate solution to this
problem as well as the discrimination between a number of possible
values of $K_{\rm A}$ might come only with future dedicated
observations, densely covering the phases of the periastron passage (the
first chance being in March 2015). In any case, it is clear that the two
RVs by \citet{SL2001} provide additional support for the 470~d period.

\begin{table}
\caption{\fotel orbital solution for the triple system \hde.}
\label{fotel}
\begin{tabular}{lllc}
\hline\hline\noalign{\smallskip}
Parameter&Unit&Value\\
\noalign{\smallskip}\hline\noalign{\smallskip}
Orbit Ba - Bb \\
\noalign{\smallskip}\hline\noalign{\smallskip}
$P_{\rm Ba-Bb}$       &(d)      &6.004865(71) \\
$T_{\rm periastr.}$   &(RJD)    &54285.15(25) \\
$T_{\rm upper\,conj.}$&(RJD)    &54285.90     \\
e                     &         &0.094(26)    \\
$\omega$              &($^\circ$)&37(15)      \\
$K_{\rm Ba}$          &(\ks)    &37.5(2.3)    \\
$f(m)$                &(\ms)    &0.0324       \\
\noalign{\smallskip}\hline\noalign{\smallskip}
Orbit A - (Ba+Bb) \\
\noalign{\smallskip}\hline\noalign{\smallskip}
$P_{\rm A-(Ba+Bb)}$   &(d)      &470.69(48) \\
$T_{\rm periastr.}$   &(RJD)    &54983.3(1.4)\\
$T_{\rm min.RV}$      &(RJD)    &55433.0    \\
$e$                   &         &0.859(20)  \\
$\omega_{\rm A}$  &($^\circ$)   &306.8(2.8) \\
$K_{\rm Ba+Bb}/K_{\rm A}$&      &1.127(44)  \\
$K_{\rm Ba+Bb}$       &(\ks)    &87(10)     \\
$K_{\rm A}$           &(\ks)    &98(11)     \\
$a \sin i$            &(AU)     &4.10       \\
rms$_{\rm Ba\,old}$&(\ks)       &17.92      \\
rms$_{\rm Ba\,new}$&(\ks)       &4.29       \\
rms$_{\rm A\,new}$ &(\ks)       &5.66       \\
\noalign{\smallskip}\hline\noalign{\smallskip}
$\gamma$           &(\ks)       &$-21.60(46)$ \\
\hline\noalign{\smallskip}
\end{tabular}\\
\footnotesize{Note: $a$ is the semi-major axis of the 470-d orbit.}
\end{table}

\begin{table*}
\caption[]{{\tt KOREL} solutions for the triple star, $P=6\fd004865$
for the short orbit fixed.}
\label{korel}
\begin{flushleft}
\begin{tabular}{llllllllll}
\hline\hline\noalign{\smallskip}
Element                       &He\,I~5876&He\,II~5411&C\,III/C\,IV region&O\,III~5592&He\,I~6678&He\,I~7065&Mean \\
\noalign{\smallskip}\hline\noalign{\smallskip}
short orbit \\
\noalign{\smallskip}\hline\noalign{\smallskip}
$T_{\rm peri.}$ (RJD$-54285$) &0.338    & 0.296    & 0.350    & 0.366    &0.356    &0.356    &0.344 $\pm$0.025 \\
$e$                           &0.1124   &0.1122    &0.1120    &0.1123    &0.1122   &0.1123   &0.1122$\pm$0.0001\\
$\omega_{\rm Ba}$ ($^\circ$)  &49.0     &51.0      &50.4      &51.0      &51.0     &49.7     &50.4  $\pm$0.8   \\
$K_{\rm Ba}$   (km\,s$^{-1}$) &35.09    &35.60     &34.50     &34.11     &33.68    &34.61    &34.60 $\pm$0.68  \\
$f(m)$ (\ms)                  &0.02538  &0.02754   &0.02507   &0.02423   &0.02328  &0.02531  &0.0251$\pm$0.0014          \\
\noalign{\smallskip}\hline\noalign{\smallskip}
long orbit \\
\noalign{\smallskip}\hline\noalign{\smallskip}
$P_{\rm long}$ (d)            &469.91   &469.90    &469.98    &469.88    &470.00   &470.00   &469.95$\pm$0.05  \\
$T_{\rm peri.}$ (RJD$-54981$) &0.89     & 0.84     & 0.85     & 0.86     &0.86     &0.86     &0.86  $\pm$0.11  \\
$e$                           &0.866    &0.862     &0.868     &0.865     &0.867    &0.864    &0.865 $\pm$0.002 \\
$\omega_{\rm Ba+Bb}$ ($^\circ$)&123.2   &123.4     &123.2     &123.5     &123.4    &123.3    &123.3 $\pm$0.12  \\
$K_{\rm Ba+Bb}$ (km\,s$^{-1}$)&82.00    &82.90     &84.30     &83.48     &84.46    &84.30    &83.6  $\pm$1.0   \\
$K_{\rm Ba+Bb}$/$K_{\rm A}$   &0.8906   &0.8985    &0.9019    &0.8915    &0.8981   &0.8978   &0.8964$\pm$0.0044\\
$K_{\rm A}$ (km\,s$^{-1}$)    &92.07    &92.26     &93.46     &93.64     &94.04    &93.90    &93.2  $\pm$0.8   \\
$a \sin i$ (AU)               &3.761    &3.837     &3.815     &3.836     &3.844    &3.878    &3.828$\pm$0.039     \\
\noalign{\smallskip}\hline
\end{tabular}
\end{flushleft}
\end{table*}

\section{RV orbital solutions with \korele}

The \fotel orbital elements were eventually used as initial values
for the solution with the spectral-disentangling programme
\korel \citep{Had04b}. Apart from the \ion{He}{I}~5876~\AA\ line,
we used five other suitable lines and spectral segments
(\ion{He}{I}~6678, \ion{He}{I}~7065, \ion{He}{II}~5411, \ion{O}{III}~5592,
and a longer segment 5670 -- 5835~\AA, which contains
\ion{C}{III}~5695, \ion{C}{IV}~5801, and \ion{C}{IV}~5812~\AA\ plus several
stronger diffuse interstellar bands) to obtain six independent \korel solutions.

In practice, we searched the parameter space and ran a number of trial
solutions for each spectral region to find out a solution with the lowest
sum of residuals. These final solutions and the resulting elements are
summarised in Table~\ref{korel} together with
their mean values. We note that \korel does not provide error
estimates so the comparison of several independent solutions gives us
some idea about the possible errors of the elements.

We carried out several attempts to detect also some weak lines of
component Bb in \korel solution but with no result. This means that we
are unable to obtain the mass ratio between components Ba and Bb and
conclude that the component Bb is much less massive than component Ba.
This is similar to HD~165246 binary, for which \citet{May2013} found a
low mass ratio of 0.17\,.

\section{Three-dimensional orbit of HD~152246~A,B}

Having resolved  the A,B pair and detected its orbital motion on the
plane of the sky opens the possibility of computing the three-dimensional
orbit of HD~152246. To this aim, we simultaneously fit the RV and
astrometric measurements using the method described in \citet{SLBM13};
we only used FEROS and BESO data. We corrected the RVs of component Ba
listed in Table~\ref{rvtab} from the orbital motion around the centre of
mass of the Ba,Bb pair using the orbital solution given in
Table~\ref{fotel}. We adopted error bars on the A and B RV components
given by the residual of the fit in Table~\ref{rvtab}, i.e.\ 5.66 and
4.29~\kms\ for component A and B, respectively. We also neglected the
astrometric displacement of B attributable to its close companion in the VLTI
data. This is a valid assumption given such displacement is likely of
the order of 0.1~mas.

\begin{figure}
\centering
\resizebox{\hsize}{!}{\includegraphics{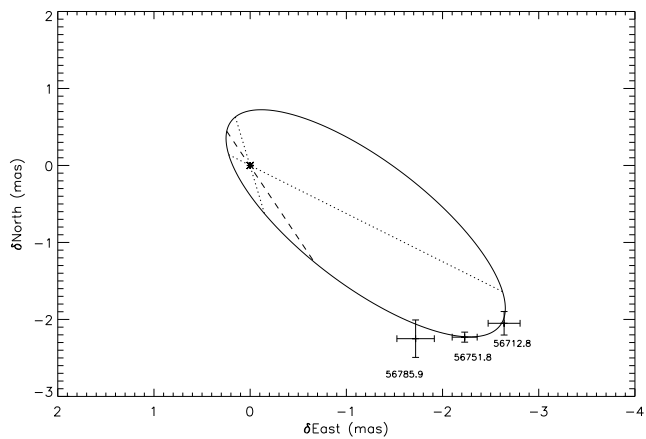}}
\caption{Astrometric orbital solution of the wide A,B pair (Table~\ref{tab: 3Dorbit}) and PIONIER interferometric measurements. The dashed line shows the line of nodes.}
\label{fig: 3Dorbit}
\end{figure}

We used the averaged \korel orbital solution as a starting point for our
fit. We fixed the distance at 1585~kpc and we worked in the systemic
velocity frame (i.e., $\gamma=0$~\kms). The obtained best-fit solution
is given in Table~\ref{tab: 3Dorbit} and shown in Figure~\ref{fig: 3Dorbit}.
As explained in \citet{SLBM13}, we estimated the uncertainties on the
best-fit parameters by Monte Carlo simulations. We randomly drew 1000
artificial RV and astrometric data sets. We preserved the observational
time sampling. The data points were taken from Gaussian distributions
centred on the best fit solution, using the observational error bars as
the Gaussian standard deviations to draw. We then ran the fitting
procedure on each of the synthetic data set and we estimated the
uncertainties on the best-fit parameters from the distribution of
retrieved parameters. The  error intervals quoted in Table~\ref{tab: 3Dorbit}
correspond to the 68\%\ confidence intervals. The intervals are taken as
symmetric if the upper and lower error bars do not differ by more than
10\%, asymmetric confidence intervals are otherwise provided.

Comparing the results of the three-dimensional orbit with the \fotel and
\korel orbital solutions (Tables~\ref{fotel} and \ref{korel}) reveal
excellent agreement. The preference for a high eccentricity value is
confirmed while we obtained slightly smaller RV semi-amplitudes.
Importantly, the orbital inclination $i$ is now constrained to
$i=112.46^{+6.98}_{-9.11}$. This allows us to derive absolute masses
for the A and B components to $20.35~\pm~1.50$ and $22.78~\pm~1.82$~\ms,
i.e.\ with a relative accuracy of 8\%.


\begin{table}
\caption{Best-fit simultaneous RV and interferometric orbital solution of the  HD~152246~A,B system.  }
\label{tab: 3Dorbit}
\begin{centering}
\begin{tabular}{lccc}
\hline\hline
Parameter        & Unit  &       A          &           B         \\
\hline
\vspace*{-3mm}\\
$P_\mathrm{out}$ &[d]    & \multicolumn{2}{c}{$470.54  \pm 0.53$} \\
$a$              &[mas]& \multicolumn{2}{c}{$2.620 \pm 0.064$} \\
$a$              &[AU]   & \multicolumn{2}{c}{$4.15 \pm 0.10$} \\
$a_j$ ($j=$A,B)  &[mas]&$1.387 \pm 0.047$ & $1.234 \pm 0.038$   \\
$a_j$ ($j=$A,B)  &[AU]   &$2.198 \pm 0.075$ & $1.955 \pm 0.060$   \\
$e$              &       & \multicolumn{2}{c}{$0.843   \pm 0.024$}\\
$\omega_{\rm A}$ &[\degr]& \multicolumn{2}{c}{$303.71  \pm 2.95$} \\
$T$              &[RJD]  & \multicolumn{2}{c}{$54983.51\pm 1.53$} \\
$\gamma$         &[\ks]  & \multicolumn{2}{c}{$-21.48  \pm 0.98$} \\

$M_{\rm A}/M_{\rm B}$    &       & \multicolumn{2}{c}{$0.893 \pm 0.038$}  \\
$i  $            &[\degr]& \multicolumn{2}{c}{$112.46^{+6.98}_{-9.11}$} \\
$\Omega$         &[\degr]& \multicolumn{2}{c}{$28.05^{+7.68}_{-6.08}$}\\
\\
$K$              &[\ks]  & $87.27 \pm10.4$  & $77.95 \pm  9.5$    \\
$M$              &[\ms]  & $20.35 \pm 1.50$ &$ 22.78 \pm  1.82$   \\
\\
$\chi^2_\mathrm{red}$ &  & \multicolumn{2}{c}{\hspace*{3mm} 0.83} \\
\vspace*{-3mm}\\
\hline
\end{tabular}\\
\end{centering}
\footnotesize{Notes: $a_{\rm A}$ and $a_{\rm B}$ denote the separations
of components A and B from the common centre of gravity
($a=a_{\rm A}+a_{\rm B}$). $\omega_{\rm A}$ refers to the direction from
the centre of mass to A at periastron.}
\end{table}

\section{Spectral disentangling and preliminary physical properties of
components A and Ba}
Keeping the mean elements of Table~\ref{korel} fixed, we used \korel
to disentangle the line profiles of components A and Ba in all six spectral
regions. Synthetic spectra were then fitted to the disentangled profiles in
order to estimate the effective temperatures $T_{\rm eff}$, gravitational
acceleration $g$, projected rotational velocity \vsin, fractional luminosity
$L_{\rm R}$, $RV$ and metallicity $Z$ of the resolved components of the system.
We attempted two independent approaches.

\subsection{{\tt TLUSTY} model atmospheres and the OSTAR grid of synthetic spectra}
One of us (JN) has developed a programme, which interpolates in a~pre-calculated
grid of synthetic stellar spectra in the effective temperature,
gravitational acceleration, and metallicity. The programme starts with an initial
estimate of all the parameters to be optimised. The interpolated spectrum of
each component is shifted in RV, broadened to the projected rotational velocity
with the programme ROTINS (written by Dr. I. Huben\'y) and multiplied by the
fractional luminosity of the modelled component. The synthetic spectrum is
compared to the disentangled one (normalised to the common continuum of the
system) and the initial parameters are optimised by minimisation of $\chi^2$
(defined by Eq.~\ref{chisquare} below) until best match is achieved:
\begin{eqnarray}
\chi^2 = \sum_{i=1}^{\rm N}w_{\rm i}\left[I_{\rm i}^{\rm disen} -
 I_{\rm i}^{\rm synt}\left(T_{\rm eff}, g, v_{\rm R}\sin i, L_{\rm R}, RV, Z\right)\right]^2
\label{chisquare}
\end{eqnarray}
Here $N$ is the number of disentangled components of the system,
$I_{i}^{\rm disen}$ the disentangled spectrum, $I_{\rm i}^{\rm synt}$
the synthetic spectrum and $w_{\rm i}$ the weight of the $i$-th
component. Fractional luminosities must satisfy the condition that
$\sum^{\rm N}_{i=1}L_{\rm R, i} = 1$ and metallicities must satisfy the
condition $Z_{i}$ = $Z_{1}$, for $i=1..N$. The $\chi^2$ is minimised
using the downhill simplex method \citep{NM65}.

All the weights of the disentangled spectra were set to one and the
interpolation was carried out in the OSTAR grid \citep{LH03} based on
{\tt TLUSTY} model atmopspheres \citep{tlusty}. Several
short segments covering the neighbourhood of the spectral lines
mentioned in Table~\ref{korel} were optimised simultaneously and for
both components at once, the \ion{C}{III}~5696~\AA\ and the
\ion{C}{IV}~5801 and 5811~\AA\ lines being treated as two separate data
segments (not as one long segment as in \korele). The results of
minimisation of Eq.~\ref{chisquare} are summarised in Table~\ref{fit}.
To estimate uncertainties of all derived parameters, we repeated the
optimisation of $\chi^{2}$ hundred times, each time starting the
optimisation from a random point. The solutions having the final
$\chi^2$ less than 1.06 times higher than the $\chi^2$ of the best
solution were used to estimate the errors. The largest source of the
error, uncertainty in the spectra (re)normalisation of the disentangled
profiles, was not taken into account\footnote{Since the disentangling in
\korel is carried out in Fourier space, it happens in almost every
practical application that the resulting disentangled spectra after
inverse Fourier transformation have slightly warped continua and need
some re-normalisation.}. This means that the errors quoted in
Table~\ref{fit} should be taken as lower limits.

On modelling side, we note that {\tt TLUSTY} models are static plane-parallel
models with full non-LTE metal line blanketing. They are suitable for
O stars that have weak stellar winds.

\begin{table}
\centering
\caption{Results of the best fit of the OSTAR grid of synthetic spectra
to the observed ones. The optimised parameters are: the effective temperature
$T_{\rm eff}$, gravitational acceleration $g$, projected rotational
velocity $v_{\rm R}\sin i$, fractional luminosity $L_{\rm R}$, $RV$, and
metallicity $Z$.}
\begin{tabular}{lr@{$\pm$}l|r@{$\pm$}l}
\hline\hline\noalign{\smallskip}
Quantity& \multicolumn{2}{c}{Component A} & \multicolumn{2}{c}{Component Ba}\\
\noalign{\smallskip}\hline\noalign{\smallskip}
$T_{\rm eff}$\,(K)&32506&23  &32544&48\\
$\log (g_{\rm [cgs]})$&3.738&0.014  &3.706&0.009\\
$v\sin i$\,(\ks)&200.4&2.9  &61.41&0.47\\
$L_{\rm R}$&0.5712&0.0021    &0.4288&0.0021\\
$RV$\,(\ks)&$-20.2$&2.6  &$-19.72$&0.36\\
$Z$\,(Z$_{\odot}$)&\multicolumn{2}{c}{1.00} &\multicolumn{2}{c}{1.00}\\
\noalign{\smallskip}\hline
\end{tabular}
\label{fit}
\end{table}

While the general fit of the disentangled by synthetic spectra resulting
from the final solution of Table~\ref{fit} was quite satisfactory,
deviations up to several per cent of the continuum level were found
in the cores of several investigated spectral lines.

There is a pronounced disagreement between the disentangled and
synthetic spectra for the \ion{C}{iii}~5696~\AA\, line.
The observed line is in emission in component~Ba.
According to the synthetic spectra, this line should be in emission only
at low \lgg. When we tried to fit this line, and simultaneously also
\ion{He}{II} and \ion{He}{I} lines, we ended with a strong disagreement
with reality. However, we note that the presence of the 5696~\AA\ emission
is common among main-sequence stars.
The case of 15~Mon (O7\,V) is known \citep{Wilson55}; see the ELODIE spectrum
\citep{Elodie}. By chance, we found that the line is in emission also in several
O8\,V stars: HD~161853 (FEROS spectrum) or in HD~49149 and HD~46966 (ELODIE
spectra).

We also note that \ion{He}{I}~6678~\AA\ line, and the two \ion{C}{IV} lines
for both, A and Ba component are stronger than suggested by the synthetic
model but this difference is only on the level of 1~\% of the continuum level
only. Once again, at least for \ion{He}{I} line, such a behaviour is not
unusual and has been observed for two other O stars, SZ Cam \citep{Lor98}, and V1331~Aql \citep{Lor2005}.

\begin{figure}
\centering
\resizebox{\hsize}{!}{\includegraphics{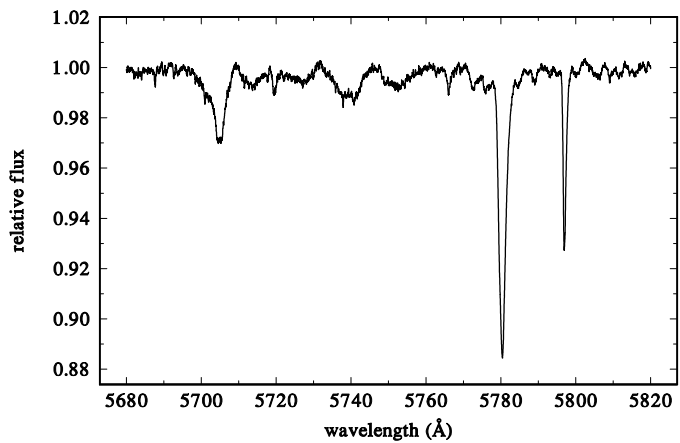}}
\caption{The disentangled spectrum of diffuse interstellar bands
in the 5680 -- 5820~\AA.}
\label{is}
\end{figure}

It would be worth of investigation whether the observed deviations are
indicative of a temperature stratification in the extended atmospheres
and/or winds around the objects in question. For those colleagues, who
study the diffuse interstellar bands, we also plot in Figure~\ref{is}
the disentangled spectrum of diffuse interstellar bands in the region
from 5680 to 5820~\AA, normalised to the joint continuum of the system.

\subsection{FASTWIND model spectra}

We also attempted to adjust the disentangled spectra with the {\tt FASTWIND} model atmosphere \citep{PUV05,fastwind} using an automatic fitting method \citep{MdKP05, TSdK11}. In contrast to {\tt TLUSTY} this programme uses spherical model atmospheres, non-LTE with line blanketing and models also stellar wind lines. On the other hand, {\tt FASTWIND} only computes H and He profiles, i.e. there is no information for the \ion{C}{iii}~5696~\AA\, line. The atmospheres are not truly hydrodynamic, however, as the programme does not solve for the wind structure but adopts a wind law. Accurate mass loss estimates from the optical require recombination lines, like H$\alpha$ or \ion{He}{II}~4686\,\AA\ which have not yet been analysed in the current investigation. The results of the fit are summarised in Table~\ref{fastwind}. A detailed fit of some spectral line was not perfect in this case either.

\begin{table}
\centering
\caption{Results of the best fit of the {\tt FASTWIND} grid of synthetic
spectra to the observed ones. Parameter $\beta$ is the usual coefficient of
the wind velocity law.}
\begin{tabular}{lc|cl}
\hline\hline\noalign{\smallskip}
Quantity& Component A & Component Ba\\
\noalign{\smallskip}\hline\noalign{\smallskip}
$L_{\rm R}$ (adopted)&0.5712  &0.4288\\
$T_{\rm eff}$\,(K)&33500 &34650\\
$\log (g_{\rm [cgs]})$&3.7  &3.7\\
$v\sin i$\,(\ks)&218 &68\\
$v_{\rm turb.}$\,(\ks)&17 &18\\
$v_\infty$\,(\ks)&$2161$&$1902$\\
$\log\dot M$ &$-6.5$&$-6.5$\\
$\beta$ &1.2&0.95\\
$R$ (\rs) &9.3 & 7.9\\
\noalign{\smallskip}\hline
\end{tabular}
\label{fastwind}
\end{table}

More generally, it is obvious that a really reliable line profile modelling
awaits further improvements, both on the theoretical and observational side.
For instance, \citet{massey} compared {\tt FASTWIND} and independent
{\tt CMFGEN} model spectra with observed spectra of 10 O stars and noted
that the surface gravities of {\tt FASTWIND} are systematically lower
by 0.12~dex compared to {\tt CMFGEN} and that both programmes have
some problems to fit the \ion{He}{I} lines for higher
luminosity stars. Concerning the observed spectra, we believe that
the spectral disentangling will provide better line profiles of
both visible stellar components after we obtain good spectra during
the periastron passage in the outer orbit.

\subsection{Preliminary physical properties of components A and Ba}
Considering the above exercises, we warn against taking the results
of Tables~\ref{fit} and \ref{fastwind} too seriously at the present stage of
investigation and postpone a more accurate determination of the
physical properties of the system for a future study. For the moment,
we consider reasonable to adopt
\tef = $33000\pm500$ and \vsin = $210\pm10$~\kms for component A, and
\tef = $33600\pm600$ and \vsin = $65\pm3$~\kms for component Ba.

\section{Physical properties of the system}
According to \citet{Sota2014} the integrated spectral type of HD~152246
is O9\,IV, and according to \citet{Mer98} $V=7$\m308, \bv = 0\m169. For
that spectral type, the expected (\bv)$_0=-0$\m30, therefore the excess
$E$(\bv) = 0\m47. HD~152246 is close to NGC~6231 (angular separation
of about 45\arcmin). \citet{Sung2013} gives for this cluster $R=3$\m20
and a distance modulus of 11.0 ($d=1585$~pc).
The systemic velocity of NGC~6231 is $-27.28 \pm 2.98$ as
obtained from 10 high-mass stars \citep{Sana08}; the scatter of individual
RVs complies with the systemic velocity of HD~152246 as given in
Table~\ref{tab: 3Dorbit} corroborating its membership in Sco OB 1 association.

Moreover, \citet{Sana05} derived basic physical properties of the massive spectroscopic and eclipsing binary CPD$-41^\circ7742$, a member of NGC\,6231 for which
they were able to derive the distance modulus of $10.92\pm0.16$, in
agreement with the above value. Applying the colour excess and a
distance modulus of 11.0 to HD~152246, we obtain $V_0=5$\m80 and $M_V=-5$\m20.

The fitting of the disentangled spectra yields the monochromatic flux ratio
between the narrow-lined component Ba and broad-lined component A as
$0.429/0.571=0.751$. This is close to the ratio 0.77 found in the $H$
band by PIONIER (Table~\ref{tab: pionier}). However, we already noted
that in this case the disentangling might not be reliable. Therefore,
we prefer the value 0.77; the corresponding magnitude difference is 0\m284.
Then the magnitude $M_{\rm V}$ of the A component is $-4$\m58, of the
B component $-4$\m30. \citet{Mart2005} has $M_V = -4$\m05 for O9\,V
and $-$5\m25 for O9\,III; the agreement with the expected magnitude
of component A (O9\,IV) is a very good one. Component Ba is then O9\,V.

The typical expected mass for HD~152246~A, based on its spectral type,
is $\sim20$~\ms\ \citep{Mart2005}, in agreement with the
derived dynamical mass. Since the expected mass for an O9\,V star is
~17.5~\ms, the secondary Bb should contribute by about 6 \ms; the mass
ratio of 0.3 between Bb and Ba is quite acceptable given the SB1 nature
of HD~152246B (the luminosity of the secondary can be neglected in the
discussion made above).

\begin{figure}
\centering
\resizebox{\hsize}{!}{\includegraphics{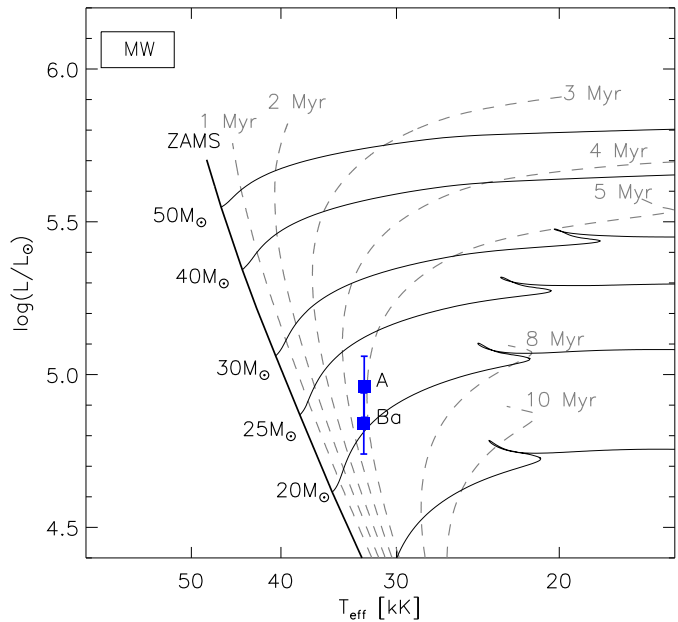}}
\caption{Position of HD~152246~A and Ba in the HRD. Evolutionary tracks
and isochrones of \citet{BdMC11} with initial rotation rate of 100~\kms\ are overlaid.}
\label{fig: hrd}
\end{figure}

\section{Discussion}

\subsection{Evolutionary status}
We used the BONNSAI\footnote{The BONNSAI web-service is available at
www.astro.uni-bonn.de/stars/bonnsai.} Bayesian tool (Schneider et al.,
submitted) to compare the observational constraints and the single massive
star evolutionary models of \citet{BdMC11}. We used a Saltpeter IMF and
flat priors for the age and initial rotational velocity.

For the A component, we provided the observed $T_\mathrm{eff}$,
$\log L_\mathrm{bol}/L_\odot$, $\log g$, $v \sin i$ and mass, together
with their uncertainty. BONNSAI is able to find a set of models that
perfectly reproduce the observed values. These correspond to initial
masses of $21.4^{+1.4}_{-0.9}$ \ms\ and age of $4.7 \pm 0.3$~Myr.

For the Ba component, we provided the same parameters with the exception
of the stellar mass, that is not directly constrained by our
observations owing to the uncertain mass of the Bb component. No model
was able to reproduce all parameters. The main issue was that our
derived $\log g$ is too low. We removed the observational constraint on
the gravity and re-ran BONNSAI. The set of models able to reproduce our
observations indicate $\log g=4.03 \pm 0.15$, a present day mass of
$18.4^{+2.0}_{-1.1}$ and an age of $4.8^{+0.4}_{-1.5}$~Myr. The
agreement of the evolutionary mass of component Ba with the expected one
based on its spectral properties is excellent. Similarly the
evolutionary ages of both A and Ba component are identical, suggesting a
coeval formation of the system. The location of both components in the
Hertzsprung-Russell diagram (HRD) is displayed in Figure~\ref{fig: hrd}.
The derived age of HD~152246 agrees with that of other high-mass
stars in NGC~6231 although the latter ones are model-dependent \citep{Sung2013}. If the
rotating models of \citet{EGE12} are adopted the age is $4.0 - 7.0$\,Myr.

\subsection{Rotational velocities and dynamical status}

Intriguingly, the very similar A and Ba components in HD~152246 presents
very different projected rotational velocities of 200 and 60~\ks.
Assuming the rotation and orbital spins of the wide binary are aligned,
we would derive very similar values of 220 and 65~\kms.

However, given the masses of the Ba, Bb system are relatively well
constrained to about 18.4 and 4.4~\ms, we can use the \fotel orbital
solution of Table~\ref{fotel} to estimate the orbital inclination of the
Ba, Bb pair. We obtain $i_\mathrm{Ba,Bb} \approx 32$\degr. Assuming this
time that rotation axis of the Ba component is aligned with the orbital
spin of the short period system, we then obtain $v_\mathrm{rot, Ba}
\approx 120$~\ks, i.e.\ still significantly lower that the one of the A
component.

BONNSAI best models indicate a probable radius of
$6.75^{+1.72}_{-1.20}$~\rs\ for the Ba component. Tidal synchronisation
of Ba in the Ba,Bb 6 d orbit would result in rotational velocity of
$57^{+14}_{-10}$~\ks. Such velocity is in agreement with the measured
projected rotational velocity and would suggest an inclination close to
90\degr\ for the spin axis of HD~152246~Ba. If this scenario is correct,
then it would mean that the spin and orbital axis are not aligned in the
Ba, Bb system. One might also speculate whether the star formation process
had not allotted similar amounts of angular momentum to both components A and B.
If so, this resulted in fast spin for A and the creation of a binary star in B.\\

\section{Conclusions}
Several conclusions can be drawn here:
\begin{enumerate}
\item[i.]  the evolutionary ages of the O stars in HD~152246 suggest a
co-eval formation of the triple system;
\item[ii.] with orbital inclination $i_\mathrm{A,B}\approx112$\degr\
and $i_\mathrm{Ba,Bb} \approx 30$ (or 150)\degr\ the two orbits are not
co-aligned;
\item[iii.] the components A and Ba likely have different rotational
velocities;
\item[iv.] the component Ba is possibly synchronised, but if this is
correct the spin and orbital axis are not aligned;
\item[v.] HD~152246~A,B pair has an eccentricity significantly higher
than expected from the period-eccentricity diagram of other O-type
systems \citep{SdMdK12}.
\end{enumerate}

It is possible that several of these properties result from the
hierarchical nature and extreme eccentricity of the system. In this
scenario, each close encounter of the Ba,Bb pair with the A component
would perturb the Ba, Bb inclination and spin alignments. Even if the A
and Ba components were born identical, the regular interactions between
the three bodies may result in different rotation velocities.
Alternatively, HD~152246 may be born as hierarchical quadruple systems
formed by two close-binaries. The Aa, Ab initial binary may have been
driven into coalescence because of stellar evolution (if the initial
period was extremely short) or  dynamical interaction with the Ba, Bb
pair. This may have lead to a catastrophic event that may have pumped up
the eccentricity of the wide system and left a rapidly rotating A
component. Such event should however have happened early in the
evolutionary life of HD~152246 as no rejuvenation effect is seen in the
HRD. Future abundance determination of the HD~152246 components may help
provide further constraints on these scenarios.

\begin{acknowledgements}
This publication is supported as a project of the Nordrhein-Westf\"{a}lische
Akademie der Wissenschaften und der K\"unste in the framework of the academy
 programme by the Federal Republic of Germany and the state Nordrhein-Westfalen.
The research of PH, PM, and JN was supported by the grant P209/10/0715 of the
 Czech Science Foundation and from the research programme MSM0021620860.
We are grateful for the help of A. Barr Dom\'{i}nguez, K. Fuhrmann, L. Kaderhandt
 and M. Ramolla during the observations and the reduction.
We thank Universidad Cat\'{o}lica del Norte in Antofagasta, Chile, for
continuous support.
The use of the NASA/ADS bibligraphical service and SIMBAD electronic database
 are gratefully acknowledged.

PIONIER is funded by the Universit\'e Joseph Fourier (UJF), the
Institut de Plan\'etologie et d'Astrophysique de Grenoble (IPAG), the
Agence Nationale pour la Recherche (ANR-06-BLAN-0421 and ANR-10-BALN-0505),
and the Institut National des Science de l'Univers (INSU PNPand PNPS).The
integrated optics beam combiner result from a collaboration between IPAG
and CEA=LETI based on CNRS R\&T funding.
\end{acknowledgements}

\end{document}